\documentclass[aps,pra,twocolumn,amsmath,amssymb]{revtex4}

\pdfoutput=1

\usepackage{graphicx}
\usepackage{dcolumn}
\usepackage{bm}

\newcommand{\ybion}{Yb${}^{+}$}

\newcommand{\ii}{\dot{\imath}}

\newcommand{\ket}[1]{\vert{#1}\rangle}
\newcommand{\bra}[1]{\langle{#1}\vert}

\newcommand{\qed}{\nobreak \ifvmode \relax \else
      \ifdim\lastskip<1.5em \hskip-\lastskip
      \hskip1.5em plus0em minus0.5em \fi \nobreak
      \vrule height0.75em width0.5em depth0.25em\fi}

\begin{document}

\title{Quantum Teleportation Between Distant Matter Qubits\footnote{Published in \textit{Science} \textbf{323}, 486 (2009)}}

\author{S. Olmschenk$^1$}
	\email{smolms@umd.edu}
\author{D. N. Matsukevich$^1$}
\author{P. Maunz$^1$}
\author{D. Hayes$^1$}
\author{L.-M. Duan$^2$}
\author{C. Monroe$^1$}
	\affiliation{${}^1$Joint Quantum Institute (JQI), University of Maryland Department of Physics and the National Institute of Standards and Technology, College Park, MD 20742, USA \\ ${}^2$FOCUS Center and Department of Physics, University of Michigan, Ann Arbor, MI 48109, USA}

\begin{abstract}
Quantum teleportation is the faithful transfer of quantum states between systems, relying on the prior establishment of entanglement and using only classical communication during the transmission.  We report teleportation of quantum information between atomic quantum memories separated by about 1 meter.  A quantum bit stored in a single trapped ytterbium ion (\ybion) is teleported to a second \ybion atom with an average fidelity of 90\% over a replete set of states.  The teleportation protocol is based on the heralded entanglement of the atoms through interference and detection of photons emitted from each atom and guided through optical fibers.  This scheme may be used for scalable quantum computation and quantum communication.
\end{abstract}

\maketitle

A defining feature of quantum physics is the inherent uncertainty of physical properties, despite the fact that we observe only definite states after a measurement.  The conventional interpretation is that the measurement process itself can irreversibly influence the quantum system under study.  The field of quantum information science makes use of this notion and frames quantum mechanics in terms of the storage, processing, and communication of information.  In particular, the back-action of measurement underlies the quantum "no cloning" theorem, which states that it is impossible to generate identical copies of an unknown quantum state~\cite{wootters:no-clone}.  Nevertheless, a quantum state can still be transferred from one system to another by the process of quantum teleportation~\cite{bennett:teleportation}.  A quantum state initially stored in system $A$ can be teleported to system $B$ by using the resource of quantum entanglement or the quantum correlation between systems that do not have well-defined individual properties.  Relaying the result of a destructive measurement of system $A$ allows the original quantum state to be recovered at system $B$ without ever having traversed the space between the systems. The ability to teleport quantum information is an essential ingredient for the long-distance quantum communication afforded by quantum repeaters~\cite{briegel:quantum_repeater} and may be a vital component to achieve the exponential processing speed-up promised by quantum computation~\cite{gottesman:qc_teleportation}.

The experimental implementation of teleportation has been accomplished in optical systems by using down-converted photons~\cite{bouwmeester:photon_teleport, boschi:photon_teleport} and squeezed light with continuous variable entanglement~\cite{furusawa:cont_var_teleport}.  Teleportation has also been accomplished between photons and a single atomic ensemble~\cite{sherson:light-ensemble_teleport, chen:light-ensemble_teleport}.  Because photons are able to carry quantum information and establish entanglement over long distances, these experiments demonstrated the nonlocal behavior of teleportation.  However, a quantum memory is required at both transmitting and receiving sites in order to scale this protocol to quantum networks and propagate quantum information over multiple nodes~\cite{duan:dlcz}. Deterministic teleportation between quantum memories has been demonstrated with trapped atomic ions in close proximity to one another, relying on the mutual Coulomb interaction~\cite{riebe:ion_teleport, barrett:ion_teleport, riebe:ion_teleport2}.  In contrast to the optical systems, these implementations feature long-lived coherences stored in good quantum memories but lack the ability to easily transmit quantum information over long distances.

\begin{figure}
	\centering
	\includegraphics[width=1.0\columnwidth,keepaspectratio]{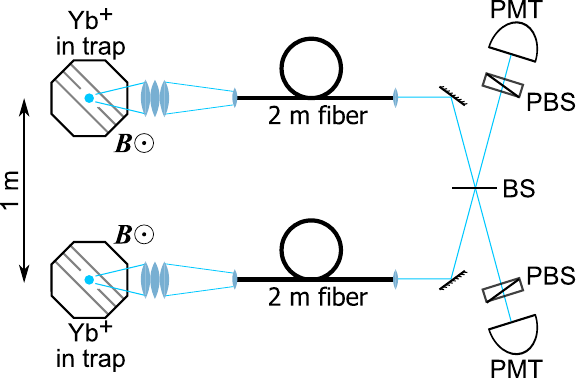}
	\caption{The experimental setup.  Two \ybion ions are trapped in independent vacuum chambers.  An externally applied magnetic field \textbf{\textit{B}} determines a quantization axis for defining the polarization of photons emitted by each atom.  Spontaneously emitted photons are collected with an objective lens, coupled into a single-mode fiber, and directed to interfere on a beamsplitter (BS).  Polarizing beamsplitters
(PBSs) filter out photons resulting from $\sigma$ decays in the atoms.  The remaining $\pi$-polarized photons are detected by single-photon counting PMTs.}
	\label{fig:teleport_arxiv_fig1}
\end{figure}

We present the implementation of a heralded teleportation protocol where the advantages from both optical systems and quantum memories are combined to teleport quantum states between two trapped ytterbium ion (\ybion) quantum bits (qubits) over a distance of about 1 m.  We fully characterized the system by performing tomography on the teleported states, enabling complete process tomography of the teleportation protocol.  The measured average teleportation fidelity of $90 \pm 2$\% (90(2)\%) over a set of mutually unbiased basis states, which is well above the 2/3 fidelity threshold that could be achieved classically, unequivocally demonstrates the quantum nature of the process~\cite{massar:teleport_threshold, enk:entangle_verify}.  Our teleportation protocol represents the implementation of a probabilistic measurement-based gate that could be used to generate entangled states for scalable quantum computation~\cite{duan:freq-qubit, vanmeter:architecture_shor}.

A schematic of the experimental setup (Fig.~\ref{fig:teleport_arxiv_fig1}) shows a single \ybion atom confined and Doppler laser-cooled in each of two nearly identical radiofrequency (rf) Paul traps, located in independent vacuum chambers~\cite{maunz:interference, moehring:ion-ion, olmschenk:state-detect, matsukevich:bell_ion}.  An ion will typically remain in the trap for several weeks.  The qubit states in each atom are chosen to be the first-order magnetic field-insensitive hyperfine ``clock'' states of the ${}^{2}S_{1/2}$ level, $\ket{F = 0, m_F = 0}$ and $\ket{F = 1, m_F = 0}$, which are separated by 12.6 GHz and defined to be $\ket{0}$ and $\ket{1}$, respectively.  In this notation, $F$ is the total angular momentum of the atom, and $m_F$ is its projection along a quantization axis defined by an external magnetic field \textbf{\textit{B}}.  The qubit exhibits coherence times observed to be greater than 2.5 s and thus serves as an excellent quantum memory~\cite{olmschenk:state-detect}.

\begin{figure*}
	\centering
	\includegraphics[width=2.0\columnwidth,keepaspectratio]{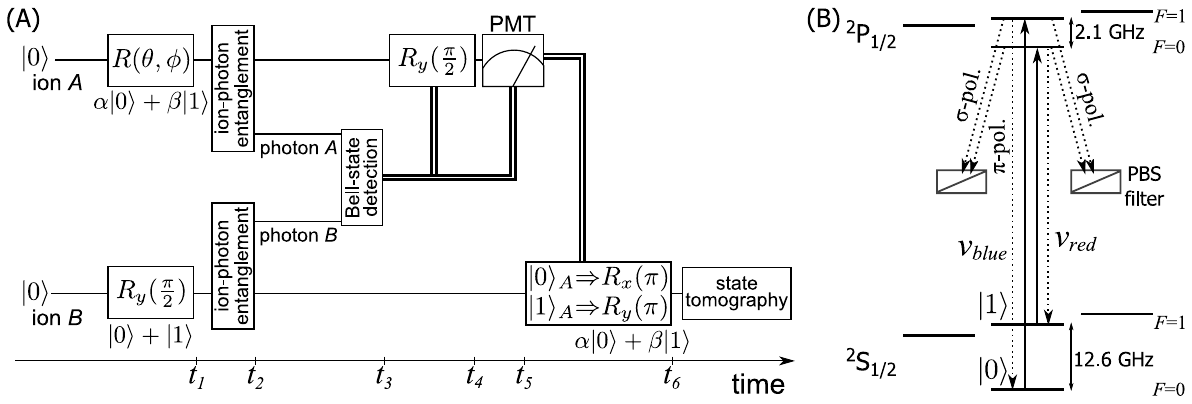}
	\caption{(\textbf{A}) Schematic of the teleportation protocol.  Each ion is first initialized to $\ket{0}$ by optical pumping.  The state to be teleported is written to ion $A$ by a microwave pulse, whereas a separate microwave pulse prepares ion $B$ in a known superposition ($t_1$).  A laser pulse excites each atom, as shown in (B).  The frequency of an emitted $\pi$-polarized photon (selected by polarization filtering) is then entangled with the hyperfine levels of the atom ($t_2$).  These two photons interfere at a BS, as illustrated in Fig.~\ref{fig:teleport_arxiv_fig1}, resulting in a coincident detection only if the photons are in the $\ket{\Psi^{-}}_{photons}$ state, which heralds the success of the ion-ion entangling gate ($t_3$).  If the gate is successful, ion $A$ is rotated by $\pi/2$ ($t_4$) and measured ($t_5$).  A microwave pulse with phase conditioned on the outcome of the measurement on ion $A$ is then applied to ion $B$ to complete the teleportation of the quantum state ($t_6$).  (\textbf{B}) Ion-photon entanglement process.  A broadband picosecond pulse with a central wavelength at 369.5 nm is used to coherently excite $\ket{0}$ and $\ket{1}$ to the ${}^{2}P_{1/2}$ level.  Because of the atomic selection rules and polarization filtering with PBSs to only observe photons from a $\pi$ decay, the coherence of the atomic states is retained.}
	\label{fig:teleport_arxiv_fig2}
\end{figure*}

For the teleportation protocol (Fig.~\ref{fig:teleport_arxiv_fig2}A), the states of the atomic qubits are initialized with a 1-$\mu$s pulse of 369.5-nm light resonant with the ${}^{2}S_{1/2} \ket{F = 1} \leftrightarrow {}^{2}P_{1/2} \ket{F = 1}$ transition that optically pumps the ions to $\ket{0}$ with probability greater than 99\%~\cite{olmschenk:state-detect}.  We can then prepare any superposition of $\ket{0}$ and $\ket{1}$ by applying a resonant microwave pulse with controlled phase and duration (0 to 16 $\mu$s) directly to one of the trap electrodes.  The quantum state to be teleported is written to ion $A$ by using this microwave pulse, which prepares ion $A$ in the state $\ket{\Psi (t_1)}_A = \alpha \ket{0}_A + \beta \ket{1}_A$.  A separate microwave pulse prepares ion $B$ in the definite state $\ket{\Psi (t_1)}_B = \ket{0}_B + \ket{1}_B$, where for simplicity we neglect normalization factors and assume ideal state evolution throughout our discussion.  After this state preparation, each ion is excited to the ${}^{2}P_{1/2}$ level with near-unit probability by an ultrafast laser pulse ($\approx$1 ps) having a linear polarization aligned parallel to the quantization axis ($\pi$-polarized) and a central wavelength of 369.5 nm.  Due to the polarization of the pulse and atomic selection rules, the broadband pulse coherently transfers $\ket{0}$ to ${}^{2}P_{1/2} \ket{F=1,m_F=0}$ and $\ket{1}$ to ${}^{2}P_{1/2} \ket{F=0,m_F=0}$ (Fig.~\ref{fig:teleport_arxiv_fig2}B)~\cite{madsen:ultrafast-rabi}.  Because the duration of this pulse is much shorter than the $\tau \approx 8$ ns natural lifetime of the ${}^{2}P_{1/2}$ level, each ion spontaneously emits a single photon while returning to the ${}^{2}S_{1/2}$ ground state~\cite{maunz:interference}.  The emitted photons at 369.5 nm can each be collected along a direction perpendicular to the quantization axis by objective lenses of numerical aperture $NA = 0.23$ and coupled into single-mode fibers.  Observation along this direction allows for polarization filtering of the emitted photons because those produced by $\pi$ and $\sigma$ transitions appear as orthogonally polarized~\cite{blinov:ion-photon}.  Considering only $\pi$ decays results in each ion being entangled with the frequency of its emitted photon such that
\begin{align}
	\label{eq:ion-photon_entangle}
	\ket{\Psi (t_2)}_A & = \alpha \ket{0}_A \ket{\nu_{blue}}_A + \beta \ket{1}_A \ket{\nu_{red}}_A \nonumber \\
	\ket{\Psi (t_2)}_B & =  			\ket{0}_B \ket{\nu_{blue}}_B + 			 \ket{1}_B \ket{\nu_{red}}_B ,
\end{align}
where $\ket{\nu_{blue}}$ and $\ket{\nu_{red}}$ are single photon states having well-resolved frequencies $\nu_{blue}$ and $\nu_{red}$, each with a bandwidth of $1/(2 \pi \tau) \approx 20$ MHz and frequency difference $\nu_{blue} - \nu_{red} = 14.7$ GHz.  The outputs of the fibers are directed to interfere at a 50:50 nonpolarizing beamsplitter, with a measured mode overlap greater than 98\%.  Because of the quantum interference of the two photons, a simultaneous detection at both output ports of the beamsplitter occurs only if the photons are in the state $\ket{\Psi^{-}}_{photons} = \ket{\nu_{blue}} \ket{\nu_{red}} - \ket{\nu_{red}} \ket{\nu_{blue}}$~\cite{hong:HOM, shih:HOM, braunstein:bs_bell_measure}, which projects the ions into the entangled state~\cite{simon:dist-entangle}:
\begin{multline}
	\label{eq:ion-ion_entangle}
	\bra{\Psi^{-} (t_3)}_{photons} \left( \ket{\Psi (t_3)}_A \otimes \ket{\Psi (t_3)}_B \right) = \\
	\ket{\Psi (t_3)}_{ions} = \alpha \ket{0}_A \ket{1}_B - \beta \ket{1}_A \ket{0}_B .
\end{multline}

A coincident detection of two photons is therefore the heralding event that announces the success
of the ion-ion entangling gate operation $\frac{1}{2} \hat{\sigma}_3^A \left( \hat{\sigma}_0^A \hat{\sigma}_0^B - \hat{\sigma}_3^A \hat{\sigma}_3^B \right)$, where $\hat{\sigma}_0^i$ is the identity and $\hat{\sigma}_3^i$ the $z$-Pauli operator acting on the $i$th qubit~\cite{duan:freq-qubit}.  In the current setup, this entangling gate only succeeds with probability $P_{gate} \approx 2.2 \times 10^{-8}$, limited by the efficiency of collecting and detecting both spontaneously emitted photons.  Therefore, the previous steps (state preparation and pulsed excitation) are repeated at a rate of 40 to 75 kHz, including intermittent cooling, until the gate operation is successful (every 12 min, on average).  Because each attempt is independent of all others, this protocol allows for a sequence of unknown and unrelated input states.  After the entanglement has been confirmed by the heralding event, another pulse of microwaves transforms the state of ion $A$ through the rotation operator $R_y(\pi/2)$, altering the state of the ions given in Eq.~\ref{eq:ion-ion_entangle} to
\begin{align}
	\label{eq:rotate_a}
		\ket{\Psi (t_4)}_{ions} = \alpha \left( \ket{0}_A + \ket{1}_A \right) & \ket{1}_B - \nonumber \\
															\beta \left( -\ket{0}_A + \ket{1}_A \right) & \ket{0}_B .
\end{align}

We then measure the state of ion $A$ with standard fluorescence techniques, by illuminating the ion with laser light at 369.5 nm, resonant with the ${}^{2}S_{1/2} \ket{F=1} \leftrightarrow {}^{2}P_{1/2} \ket{F=0}$ transition.  If the ion is in the state $\ket{1}$, it scatters many photons, whereas if the ion is in the state $\ket{0}$ the light is off-resonance and almost no photons are scattered.  By detecting the fluorescence of the
atom with a single-photon counting photomultiplier tube (PMT), we discriminate between $\ket{0}$ and $\ket{1}$ with an error of about 2\%~\cite{olmschenk:state-detect}.

Measuring ion $A$ projects ion $B$ into one of the two states:
\begin{align}
	\label{eq:measure_ion_a}
	\ket{\Psi (t_5)}_B & = \alpha \ket{1}_B + \beta \ket{0}_B && \text{(if measured $\ket{0}_A$)} \nonumber \\
	\ket{\Psi (t_5)}_B & = \alpha \ket{1}_B - \beta \ket{0}_B && \text{(if measured $\ket{1}_A$)} .
\end{align}

The result of the measurement on ion $A$ is relayed through a classical communication channel and used to determine the necessary phase of a conditional microwave $\pi$ pulse applied to ion $B$ to recover the state initially written to ion $A$; measuring $\ket{0}_A$ requires the rotation $R_x(\pi)$, whereas $\ket{1}_A$ demands $R_y(\pi)$.  Afterward, the state of ion $B$ is ideally $\ket{\Psi (t_6)}_B = \alpha \ket{0}_B + \beta \ket{1}_B$, which completes the teleportation of the quantum state between the two distant matter qubits.

The teleportation protocol we present differs from the original proposal~\cite{bennett:teleportation} in that we use four qubits (two atoms and two photons) rather than three, and our implementation is intrinsically probabilistic because the two-photon Bell states are not all deterministically distinguishable~\cite{bouwmeester:photon_teleport, braunstein:bs_bell_measure, simon:dist-entangle}.  Nevertheless, the heralding event of the two-photon coincident detection still allows our teleportation protocol to succeed without postselection~\cite{enk:entangle_verify}. In addition, establishing the quantum channel between the (atomic) quantum memories with photons and entanglement swapping allows the atoms to be separated by a large distance from the outset.

A successful implementation of this teleportation protocol requires the transmission of two classical bits of information:  one to announce the success of the entangling gate and another to determine the proper final rotation to recover the teleported state at ion $B$.  Although these classical bits do not contain any information about $\alpha$ or $\beta$, in the absence of this classical information ion $B$ is left in a mixed state (Eq.~\ref{eq:measure_ion_a}), and the protocol fails.  The required classical communication assures that no information is transferred superluminally~\cite{bennett:teleportation}.

We evaluate the teleportation protocol by performing state tomography on each teleported state.  The tomographic reconstruction of the single-qubit density matrix can be completed by measuring the state in three mutually unbiased measurement bases.  Because measurement of the
ion occurs via the aforementioned state fluorescence technique, measurement in the remaining two bases requires an additional microwave pulse before detection; we define the rotation \{$R_y(\pi/2)$, $R_x(\pi/2)$, $R(0)$\} before detection to correspond to measurement in the basis \{$x$, $y$, $z$\}.  The single-qubit density matrix is then reconstructed from these measurements with use of a simple analytical expression~\cite{altepeter:tomography}.

\begin{figure*}
	\centering
	\includegraphics[width=2.0\columnwidth,keepaspectratio]{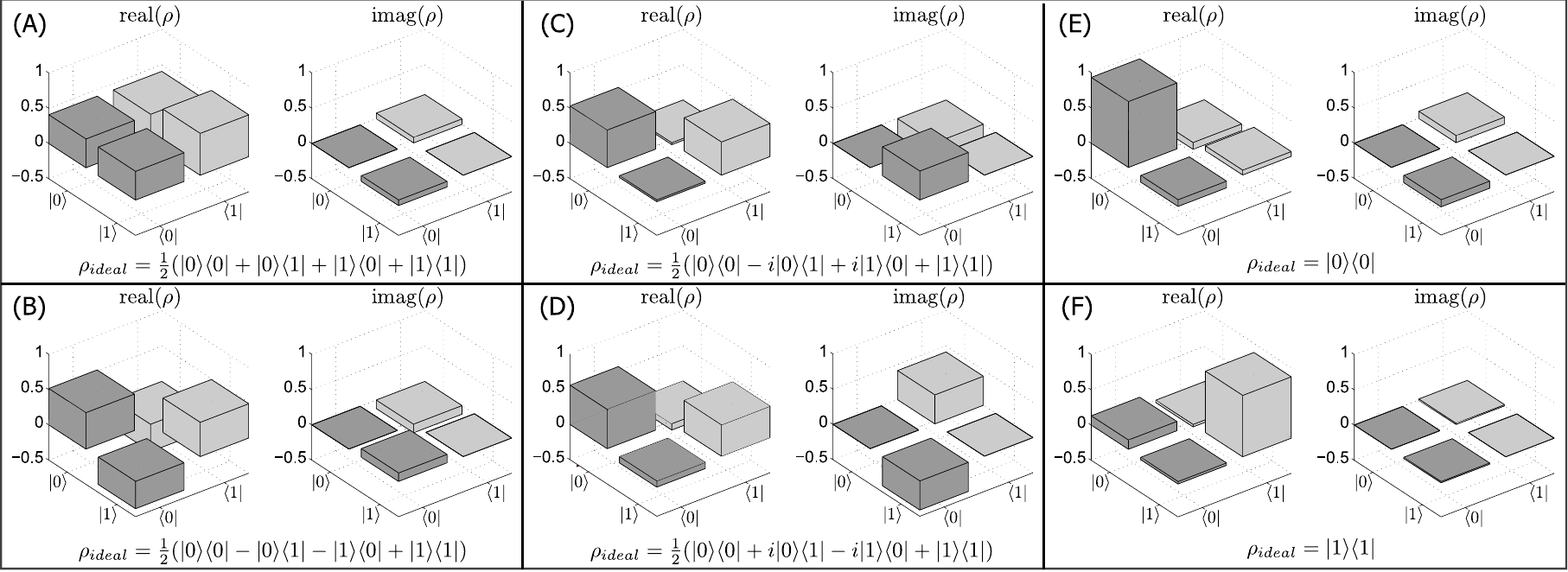}
	\caption{Tomography of the teleported quantum states.  The reconstructed density matrices, $\rho$, for the six unbiased basis states teleported from ion $A$ to ion $B$:  (\textbf{A}) $\ket{\Psi_{ideal}} = \ket{0} + \ket{1}$ teleported with fidelity $f = 0.91(3)$, (\textbf{B}) $\ket{\Psi_{ideal}} = \ket{0} - \ket{1}$ teleported with fidelity $f = 0.88(4)$, (\textbf{C}) $\ket{\Psi_{ideal}} = \ket{0} + \ii \ket{1}$ teleported with fidelity $f = 0.92(4)$, (\textbf{D}) $\ket{\Psi_{ideal}} = \ket{0} - \ii \ket{1}$ teleported with fidelity $f = 0.91(4)$, (\textbf{E}) $\ket{\Psi_{ideal}} = \ket{0}$ teleported with fidelity $f = 0.93(4)$, and (\textbf{F}) $\ket{\Psi_{ideal}} = \ket{1}$ teleported with fidelity $f = 0.88(4)$.  These measurements yield an average teleportation fidelity $\bar{f} = 0.90(2)$, where we have defined the fidelity as the overlap of the ideal teleported state with the measured density matrix, $f = \bra{\Psi_{ideal}} \rho \ket{\Psi_{ideal}}$.  The data shown comprise a total of 1285 events in 253 hours.}
	\label{fig:teleport_arxiv_fig3}
\end{figure*}

We teleport and perform tomography on the set of six mutually unbiased basis states $\ket{\Psi_{ideal}} \in$ \{ $\ket{0} + \ket{1}$, $\ket{0} - \ket{1}$, $\ket{0} + \ii \ket{1}$, $\ket{0} - \ii \ket{1}$, $\ket{0}$, $\ket{1}$ \}.  The reconstructed density matrix, $\rho$, for each of these teleported states is shown in Fig.~\ref{fig:teleport_arxiv_fig3}.  The fidelity of the teleportation, defined as the overlap of the ideal teleported state and the measured density matrix $f = \bra{\Psi_{ideal}} \rho \ket{\Psi_{ideal}}$, for this set of states is measured to be $f =$  \{0.91(3), 0.88(4), 0.92(4), 0.91(4), 0.93(4), 0.88(4)\}, yielding an average teleportation fidelity $\bar{f} = 0.90(2)$.  The experimental teleportation fidelities surpass the maximum value of $2/3$ that is achievable by classical means, explicitly demonstrating the quantum nature of the process~\cite{massar:teleport_threshold, enk:entangle_verify}.

\begin{figure}
	\centering
	\includegraphics[width=1.0\columnwidth,keepaspectratio]{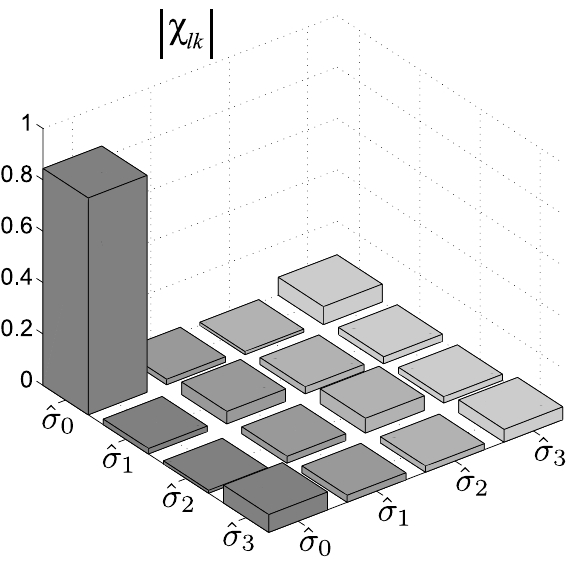}
	\caption{Absolute value of the components of the reconstructed process matrix, $\left| \chi_{lk} \right|$, with $l$, $k$ = 0, 1, 2, and 3.  The state tomography of the six mutually unbiased basis states teleported between the two ions, displayed in Fig.~\ref{fig:teleport_arxiv_fig3}, enables process tomography of the teleportation protocol by a maximum likelihood method.  The operators $\hat{\sigma}_i$ are the identity ($i = 0$) and the $x$-, $y$-, and $z$-Pauli matrices ($i = 1$, 2, and 3).  As intended, the dominant component of $\chi$ is the contribution of the
identity operation, yielding an overall process fidelity $f_{process} = \text{tr} \left( \chi_{ideal} \chi \right) = 0.84(2)$, consistent with the average fidelity cited above.}
	\label{fig:teleport_arxiv_fig4}
\end{figure}

The reconstructed density matrices also facilitate full characterization of the teleportation protocol by quantum process tomography.  We can completely describe the effect of the teleportation protocol on an input state $\rho_{in}$ by determining the process matrix $\chi$, defined by $\rho = \sum_{l,k = 0}^3 \chi_{lk} \hat{\sigma}_l \rho_{in} \hat{\sigma}_k$, where to evaluate our process we take $\rho_{in} = \ket{\Psi_{ideal}} \bra{\Psi_{ideal}}$.  The ideal process matrix, $\chi_{ideal}$, has only one nonzero component, $\left( \chi_{ideal} \right)_{00} = 1$, meaning the input state is faithfully teleported.  We experimentally determine the process matrix $\chi$ (Fig.~\ref{fig:teleport_arxiv_fig4}) by using a maximum likelihood method~\cite{obrien:process_tomography} and calculate the process fidelity to be $f_{process} = \text{tr} \left( \chi_{ideal} \chi \right) = 0.84(2)$.  Given that the relation between the average fidelity and process fidelity is $f_{process} = \left( 3 \bar{f} - 1 \right)/2$, this is consistent with the average fidelity found above~\cite{horodecki:teleport_channel}.

The deviation from unit average fidelity is consistent with known experimental errors.  The primary sources that reduce the average fidelity are imperfect state detection (3.5\%), photon mode mismatch at the 50:50 beamsplitter (4\%), and polarization mixing resulting from the nonzero numerical aperture of the objective lens and from misalignment with respect to the magnetic field (2\%).  Other sources, including incomplete state preparation, pulsed excitation to the wrong atomic state, dark counts of the PMT leading to false coincidence events, photon polarization rotation while traversing the optical fiber, and multiple excitation resulting from pulsed laser light leakage, are each expected to contribute to the error by much less than 1\%.  Residual micromotion at the rf-drive frequency of the ion trap, which alters the spectrum of the emitted photons and degrades the quantum interference, reduces the average fidelity by less than 1\%.

The entangling gate central to this teleportation protocol is a heralded, probabilistic process.  The net probability for coincident detection
of two emitted photons is given by $P_{gate} = \left( p_{Bell} \right) \left[ p_{\pi} \eta T_{fiber} T_{optics} \xi \left( \Delta \Omega / 4 \pi \right) \right]^2 \approx 2.2 \times 10^{-8}$, where $p_{Bell} = 0.25$ accounts for the detection of only one out of the four possible Bell states;
$p_{\pi} = 0.5$ is the fraction of photons with the correct polarization (half are filtered out as being produced by $\sigma$ decays); $\eta = 0.15$ is the quantum efficiency of each PMT; $T_{fiber} = 0.2$ is the coupling and transmission of each photon through the single-mode optical fiber; $T_{optics} = 0.95$ is the transmission of each photon through the other optical components; $\xi = 1 - 0.005 = 0.995$, where 0.005 is the branching ratio into the ${}^{2}D_{3/2}$ level; and $\Delta \Omega / 4 \pi = 0.02$ is the solid angle of light collection.  The attempt rate of 75 kHz is currently limited by the time of the state preparation microwave pulse, resulting in about one successful teleportation event every 12 min.  However, the expression for $P_{gate}$ reveals multiple ways to substantially increase the success rate.  The most dramatic increase would be achieved by increasing the effective solid angle of collection, which, for instance, could be accomplished by surrounding each ion with an optical cavity.  Although improvements that increase the success probability of the gate operation can enhance scalability, even with a low success probability this gate can still be scaled to more complex systems~\cite{duan:freq-qubit}.

The fidelity obtained in the current experiment is evidence of the excellent coherence properties of the photonic frequency qubit and the ``clock'' state atomic qubit.  Together, these complementary qubits provide a robust system for applications in quantum information.  The teleportation scheme demonstrated here could be used as the elementary constituent of a quantum repeater.  Moreover, the entangling gate implemented in this protocol may be used for scalable measurement-based quantum computation.

\begin{acknowledgments}
This work is supported by the Intelligence Advanced Research Projects Activity (IARPA) under Army Research Office contract, the NSF Physics at the Information Frontier program, and the NSF Physics Frontier Center at JQI.
\end{acknowledgments}


\begin{thebibliography}{30}
\expandafter\ifx\csname natexlab\endcsname\relax\def\natexlab#1{#1}\fi
\expandafter\ifx\csname bibnamefont\endcsname\relax
  \def\bibnamefont#1{#1}\fi
\expandafter\ifx\csname bibfnamefont\endcsname\relax
  \def\bibfnamefont#1{#1}\fi
\expandafter\ifx\csname citenamefont\endcsname\relax
  \def\citenamefont#1{#1}\fi
\expandafter\ifx\csname url\endcsname\relax
  \def\url#1{\texttt{#1}}\fi
\expandafter\ifx\csname urlprefix\endcsname\relax\def\urlprefix{URL }\fi
\providecommand{\bibinfo}[2]{#2}
\providecommand{\eprint}[2][]{\url{#2}}

\bibitem[{\citenamefont{Wootters and Zurek}(1982)}]{wootters:no-clone}
\bibinfo{author}{\bibfnamefont{W.~K.} \bibnamefont{Wootters}} \bibnamefont{and}
  \bibinfo{author}{\bibfnamefont{W.~H.} \bibnamefont{Zurek}},
  \bibinfo{journal}{Nature} \textbf{\bibinfo{volume}{299}},
  \bibinfo{pages}{802} (\bibinfo{year}{1982}).

\bibitem[{\citenamefont{Bennett et~al.}(1993)\citenamefont{Bennett, Brassard,
  {Cr\'{e}peau}, Jozsa, Peres, and Wootters}}]{bennett:teleportation}
\bibinfo{author}{\bibfnamefont{C.~H.} \bibnamefont{Bennett}},
  \bibinfo{author}{\bibfnamefont{G.}~\bibnamefont{Brassard}},
  \bibinfo{author}{\bibfnamefont{C.}~\bibnamefont{{Cr\'{e}peau}}},
  \bibinfo{author}{\bibfnamefont{R.}~\bibnamefont{Jozsa}},
  \bibinfo{author}{\bibfnamefont{A.}~\bibnamefont{Peres}}, \bibnamefont{and}
  \bibinfo{author}{\bibfnamefont{W.~K.} \bibnamefont{Wootters}},
  \bibinfo{journal}{Phys. Rev. Lett.} \textbf{\bibinfo{volume}{70}},
  \bibinfo{pages}{1895} (\bibinfo{year}{1993}).

\bibitem[{\citenamefont{Briegel et~al.}(1998)\citenamefont{Briegel, {D\"{u}r},
  Cirac, and Zoller}}]{briegel:quantum_repeater}
\bibinfo{author}{\bibfnamefont{H.-J.} \bibnamefont{Briegel}},
  \bibinfo{author}{\bibfnamefont{W.}~\bibnamefont{{D\"{u}r}}},
  \bibinfo{author}{\bibfnamefont{J.~I.} \bibnamefont{Cirac}}, \bibnamefont{and}
  \bibinfo{author}{\bibfnamefont{P.}~\bibnamefont{Zoller}},
  \bibinfo{journal}{Phys. Rev. Lett.} \textbf{\bibinfo{volume}{81}},
  \bibinfo{pages}{5932} (\bibinfo{year}{1998}).

\bibitem[{\citenamefont{Gottesman and
  Chuang}(1999)}]{gottesman:qc_teleportation}
\bibinfo{author}{\bibfnamefont{D.}~\bibnamefont{Gottesman}} \bibnamefont{and}
  \bibinfo{author}{\bibfnamefont{I.~L.} \bibnamefont{Chuang}},
  \bibinfo{journal}{Nature} \textbf{\bibinfo{volume}{402}},
  \bibinfo{pages}{390} (\bibinfo{year}{1999}).

\bibitem[{\citenamefont{Bouwmeester et~al.}(1997)\citenamefont{Bouwmeester,
  Pan, Mattle, Eibl, Weinfurter, and Zeilinger}}]{bouwmeester:photon_teleport}
\bibinfo{author}{\bibfnamefont{D.}~\bibnamefont{Bouwmeester}},
  \bibinfo{author}{\bibfnamefont{J.-W.} \bibnamefont{Pan}},
  \bibinfo{author}{\bibfnamefont{K.}~\bibnamefont{Mattle}},
  \bibinfo{author}{\bibfnamefont{M.}~\bibnamefont{Eibl}},
  \bibinfo{author}{\bibfnamefont{H.}~\bibnamefont{Weinfurter}},
  \bibnamefont{and}
  \bibinfo{author}{\bibfnamefont{A.}~\bibnamefont{Zeilinger}},
  \bibinfo{journal}{Nature} \textbf{\bibinfo{volume}{390}},
  \bibinfo{pages}{575} (\bibinfo{year}{1997}).

\bibitem[{\citenamefont{Boschi et~al.}(1998)\citenamefont{Boschi, Branca, {De
  Martini}, Hardy, and Popescu}}]{boschi:photon_teleport}
\bibinfo{author}{\bibfnamefont{D.}~\bibnamefont{Boschi}},
  \bibinfo{author}{\bibfnamefont{S.}~\bibnamefont{Branca}},
  \bibinfo{author}{\bibfnamefont{F.}~\bibnamefont{{De Martini}}},
  \bibinfo{author}{\bibfnamefont{L.}~\bibnamefont{Hardy}}, \bibnamefont{and}
  \bibinfo{author}{\bibfnamefont{S.}~\bibnamefont{Popescu}},
  \bibinfo{journal}{Phys. Rev. Lett.} \textbf{\bibinfo{volume}{80}},
  \bibinfo{pages}{1121} (\bibinfo{year}{1998}).

\bibitem[{\citenamefont{Furusawa et~al.}(1998)\citenamefont{Furusawa,
  {S{\o}rensen}, Braunstein, Fuchs, Kimble, and
  Polzik}}]{furusawa:cont_var_teleport}
\bibinfo{author}{\bibfnamefont{A.}~\bibnamefont{Furusawa}},
  \bibinfo{author}{\bibfnamefont{J.~L.} \bibnamefont{{S{\o}rensen}}},
  \bibinfo{author}{\bibfnamefont{S.~L.} \bibnamefont{Braunstein}},
  \bibinfo{author}{\bibfnamefont{C.~A.} \bibnamefont{Fuchs}},
  \bibinfo{author}{\bibfnamefont{H.~J.} \bibnamefont{Kimble}},
  \bibnamefont{and} \bibinfo{author}{\bibfnamefont{E.~S.}
  \bibnamefont{Polzik}}, \bibinfo{journal}{Science}
  \textbf{\bibinfo{volume}{282}}, \bibinfo{pages}{706} (\bibinfo{year}{1998}).

\bibitem[{\citenamefont{Sherson et~al.}(2006)\citenamefont{Sherson, Krauter,
  Olsson, Julsgaard, Hammerer, Cirac, and
  Polzik}}]{sherson:light-ensemble_teleport}
\bibinfo{author}{\bibfnamefont{J.~F.} \bibnamefont{Sherson}},
  \bibinfo{author}{\bibfnamefont{H.}~\bibnamefont{Krauter}},
  \bibinfo{author}{\bibfnamefont{R.~K.} \bibnamefont{Olsson}},
  \bibinfo{author}{\bibfnamefont{B.}~\bibnamefont{Julsgaard}},
  \bibinfo{author}{\bibfnamefont{K.}~\bibnamefont{Hammerer}},
  \bibinfo{author}{\bibfnamefont{I.}~\bibnamefont{Cirac}}, \bibnamefont{and}
  \bibinfo{author}{\bibfnamefont{E.~S.} \bibnamefont{Polzik}},
  \bibinfo{journal}{Nature} \textbf{\bibinfo{volume}{443}},
  \bibinfo{pages}{557} (\bibinfo{year}{2006}).

\bibitem[{\citenamefont{Chen et~al.}(2008)\citenamefont{Chen, Chen, Yuan, Zhao,
  Chuu, Schmiedmayer, and Pan}}]{chen:light-ensemble_teleport}
\bibinfo{author}{\bibfnamefont{Y.-A.} \bibnamefont{Chen}},
  \bibinfo{author}{\bibfnamefont{S.}~\bibnamefont{Chen}},
  \bibinfo{author}{\bibfnamefont{Z.-S.} \bibnamefont{Yuan}},
  \bibinfo{author}{\bibfnamefont{B.}~\bibnamefont{Zhao}},
  \bibinfo{author}{\bibfnamefont{C.-S.} \bibnamefont{Chuu}},
  \bibinfo{author}{\bibfnamefont{J.}~\bibnamefont{Schmiedmayer}},
  \bibnamefont{and} \bibinfo{author}{\bibfnamefont{J.-W.} \bibnamefont{Pan}},
  \bibinfo{journal}{Nature Physics} \textbf{\bibinfo{volume}{4}},
  \bibinfo{pages}{103} (\bibinfo{year}{2008}).

\bibitem[{\citenamefont{Duan et~al.}(2001)\citenamefont{Duan, Lukin, Cirac, and
  Zoller}}]{duan:dlcz}
\bibinfo{author}{\bibfnamefont{L.-M.} \bibnamefont{Duan}},
  \bibinfo{author}{\bibfnamefont{M.~D.} \bibnamefont{Lukin}},
  \bibinfo{author}{\bibfnamefont{J.~I.} \bibnamefont{Cirac}}, \bibnamefont{and}
  \bibinfo{author}{\bibfnamefont{P.}~\bibnamefont{Zoller}},
  \bibinfo{journal}{Nature} \textbf{\bibinfo{volume}{414}},
  \bibinfo{pages}{413} (\bibinfo{year}{2001}).

\bibitem[{\citenamefont{Riebe et~al.}(2004)\citenamefont{Riebe, {H\"{a}ffner},
  Roos, {H\"{a}nsel}, Benhelm, Lancaster, {K\"{o}rber}, Becher,
  {Schmidt-Kaler}, James et~al.}}]{riebe:ion_teleport}
\bibinfo{author}{\bibfnamefont{M.}~\bibnamefont{Riebe}},
  \bibinfo{author}{\bibfnamefont{H.}~\bibnamefont{{H\"{a}ffner}}},
  \bibinfo{author}{\bibfnamefont{C.~F.} \bibnamefont{Roos}},
  \bibinfo{author}{\bibfnamefont{W.}~\bibnamefont{{H\"{a}nsel}}},
  \bibinfo{author}{\bibfnamefont{J.}~\bibnamefont{Benhelm}},
  \bibinfo{author}{\bibfnamefont{G.~P.~T.} \bibnamefont{Lancaster}},
  \bibinfo{author}{\bibfnamefont{T.~W.} \bibnamefont{{K\"{o}rber}}},
  \bibinfo{author}{\bibfnamefont{C.}~\bibnamefont{Becher}},
  \bibinfo{author}{\bibfnamefont{F.}~\bibnamefont{{Schmidt-Kaler}}},
  \bibinfo{author}{\bibfnamefont{D.~F.~V.} \bibnamefont{James}},
  \bibnamefont{et~al.}, \bibinfo{journal}{Nature}
  \textbf{\bibinfo{volume}{429}}, \bibinfo{pages}{734} (\bibinfo{year}{2004}).

\bibitem[{\citenamefont{Barrett et~al.}(2004)\citenamefont{Barrett, Chiaverini,
  Schaetz, Britton, Itano, Jost, Knill, Langer, Leibfried, Ozeri
  et~al.}}]{barrett:ion_teleport}
\bibinfo{author}{\bibfnamefont{M.~D.} \bibnamefont{Barrett}},
  \bibinfo{author}{\bibfnamefont{J.}~\bibnamefont{Chiaverini}},
  \bibinfo{author}{\bibfnamefont{T.}~\bibnamefont{Schaetz}},
  \bibinfo{author}{\bibfnamefont{J.}~\bibnamefont{Britton}},
  \bibinfo{author}{\bibfnamefont{W.~M.} \bibnamefont{Itano}},
  \bibinfo{author}{\bibfnamefont{J.~D.} \bibnamefont{Jost}},
  \bibinfo{author}{\bibfnamefont{E.}~\bibnamefont{Knill}},
  \bibinfo{author}{\bibfnamefont{C.}~\bibnamefont{Langer}},
  \bibinfo{author}{\bibfnamefont{D.}~\bibnamefont{Leibfried}},
  \bibinfo{author}{\bibfnamefont{R.}~\bibnamefont{Ozeri}},
  \bibnamefont{et~al.}, \bibinfo{journal}{Nature}
  \textbf{\bibinfo{volume}{429}}, \bibinfo{pages}{737} (\bibinfo{year}{2004}).

\bibitem[{\citenamefont{Riebe et~al.}(2007)\citenamefont{Riebe, Chwalla,
  Benhelm, {H\"{a}ffner}, {H\"{a}nsel}, Roos, and Blatt}}]{riebe:ion_teleport2}
\bibinfo{author}{\bibfnamefont{M.}~\bibnamefont{Riebe}},
  \bibinfo{author}{\bibfnamefont{M.}~\bibnamefont{Chwalla}},
  \bibinfo{author}{\bibfnamefont{J.}~\bibnamefont{Benhelm}},
  \bibinfo{author}{\bibfnamefont{H.}~\bibnamefont{{H\"{a}ffner}}},
  \bibinfo{author}{\bibfnamefont{W.}~\bibnamefont{{H\"{a}nsel}}},
  \bibinfo{author}{\bibfnamefont{C.~F.} \bibnamefont{Roos}}, \bibnamefont{and}
  \bibinfo{author}{\bibfnamefont{R.}~\bibnamefont{Blatt}},
  \bibinfo{journal}{New J. Phys.} \textbf{\bibinfo{volume}{9}},
  \bibinfo{pages}{211} (\bibinfo{year}{2007}).

\bibitem[{\citenamefont{Massar and Popescu}(1995)}]{massar:teleport_threshold}
\bibinfo{author}{\bibfnamefont{S.}~\bibnamefont{Massar}} \bibnamefont{and}
  \bibinfo{author}{\bibfnamefont{S.}~\bibnamefont{Popescu}},
  \bibinfo{journal}{Phys. Rev. Lett.} \textbf{\bibinfo{volume}{74}},
  \bibinfo{pages}{1259} (\bibinfo{year}{1995}).

\bibitem[{\citenamefont{{van Enk} et~al.}(2007)\citenamefont{{van Enk},
  {L\"{u}tkenhaus}, and Kimble}}]{enk:entangle_verify}
\bibinfo{author}{\bibfnamefont{S.~J.} \bibnamefont{{van Enk}}},
  \bibinfo{author}{\bibfnamefont{N.}~\bibnamefont{{L\"{u}tkenhaus}}},
  \bibnamefont{and} \bibinfo{author}{\bibfnamefont{H.~J.}
  \bibnamefont{Kimble}}, \bibinfo{journal}{Phys. Rev. A}
  \textbf{\bibinfo{volume}{75}}, \bibinfo{pages}{052318}
  (\bibinfo{year}{2007}).

\bibitem[{\citenamefont{Duan et~al.}(2006)\citenamefont{Duan, Madsen, Moehring,
  Maunz, {Kohn, Jr.}, and Monroe}}]{duan:freq-qubit}
\bibinfo{author}{\bibfnamefont{L.-M.} \bibnamefont{Duan}},
  \bibinfo{author}{\bibfnamefont{M.~J.} \bibnamefont{Madsen}},
  \bibinfo{author}{\bibfnamefont{D.~L.} \bibnamefont{Moehring}},
  \bibinfo{author}{\bibfnamefont{P.}~\bibnamefont{Maunz}},
  \bibinfo{author}{\bibfnamefont{R.~N.} \bibnamefont{{Kohn, Jr.}}},
  \bibnamefont{and} \bibinfo{author}{\bibfnamefont{C.}~\bibnamefont{Monroe}},
  \bibinfo{journal}{Phys. Rev. A} \textbf{\bibinfo{volume}{73}},
  \bibinfo{pages}{062324} (\bibinfo{year}{2006}).

\bibitem[{\citenamefont{{Van Meter} et~al.}(2005)\citenamefont{{Van Meter},
  Itoh, and Ladd}}]{vanmeter:architecture_shor}
\bibinfo{author}{\bibfnamefont{R.}~\bibnamefont{{Van Meter}}},
  \bibinfo{author}{\bibfnamefont{K.~M.} \bibnamefont{Itoh}}, \bibnamefont{and}
  \bibinfo{author}{\bibfnamefont{T.~D.} \bibnamefont{Ladd}}
  (\bibinfo{year}{2005}), \bibinfo{note}{arXiv:quant-ph/0507023}.

\bibitem[{\citenamefont{Maunz et~al.}(2007)\citenamefont{Maunz, Moehring,
  Olmschenk, Younge, Matsukevich, and Monroe}}]{maunz:interference}
\bibinfo{author}{\bibfnamefont{P.}~\bibnamefont{Maunz}},
  \bibinfo{author}{\bibfnamefont{D.~L.} \bibnamefont{Moehring}},
  \bibinfo{author}{\bibfnamefont{S.}~\bibnamefont{Olmschenk}},
  \bibinfo{author}{\bibfnamefont{K.~C.} \bibnamefont{Younge}},
  \bibinfo{author}{\bibfnamefont{D.~N.} \bibnamefont{Matsukevich}},
  \bibnamefont{and} \bibinfo{author}{\bibfnamefont{C.}~\bibnamefont{Monroe}},
  \bibinfo{journal}{Nature Physics} \textbf{\bibinfo{volume}{3}},
  \bibinfo{pages}{538} (\bibinfo{year}{2007}).

\bibitem[{\citenamefont{Moehring et~al.}(2007)\citenamefont{Moehring, Maunz,
  Olmschenk, Younge, Matsukevich, Duan, and Monroe}}]{moehring:ion-ion}
\bibinfo{author}{\bibfnamefont{D.~L.} \bibnamefont{Moehring}},
  \bibinfo{author}{\bibfnamefont{P.}~\bibnamefont{Maunz}},
  \bibinfo{author}{\bibfnamefont{S.}~\bibnamefont{Olmschenk}},
  \bibinfo{author}{\bibfnamefont{K.~C.} \bibnamefont{Younge}},
  \bibinfo{author}{\bibfnamefont{D.~N.} \bibnamefont{Matsukevich}},
  \bibinfo{author}{\bibfnamefont{L.-M.} \bibnamefont{Duan}}, \bibnamefont{and}
  \bibinfo{author}{\bibfnamefont{C.}~\bibnamefont{Monroe}},
  \bibinfo{journal}{Nature} \textbf{\bibinfo{volume}{449}}, \bibinfo{pages}{68}
  (\bibinfo{year}{2007}).

\bibitem[{\citenamefont{Olmschenk et~al.}(2007)\citenamefont{Olmschenk, Younge,
  Moehring, Matsukevich, Maunz, and Monroe}}]{olmschenk:state-detect}
\bibinfo{author}{\bibfnamefont{S.}~\bibnamefont{Olmschenk}},
  \bibinfo{author}{\bibfnamefont{K.~C.} \bibnamefont{Younge}},
  \bibinfo{author}{\bibfnamefont{D.~L.} \bibnamefont{Moehring}},
  \bibinfo{author}{\bibfnamefont{D.~N.} \bibnamefont{Matsukevich}},
  \bibinfo{author}{\bibfnamefont{P.}~\bibnamefont{Maunz}}, \bibnamefont{and}
  \bibinfo{author}{\bibfnamefont{C.}~\bibnamefont{Monroe}},
  \bibinfo{journal}{Phys. Rev. A} \textbf{\bibinfo{volume}{76}},
  \bibinfo{pages}{052314} (\bibinfo{year}{2007}).

\bibitem[{\citenamefont{Matsukevich et~al.}(2008)\citenamefont{Matsukevich,
  Maunz, Moehring, Olmschenk, and Monroe}}]{matsukevich:bell_ion}
\bibinfo{author}{\bibfnamefont{D.~N.} \bibnamefont{Matsukevich}},
  \bibinfo{author}{\bibfnamefont{P.}~\bibnamefont{Maunz}},
  \bibinfo{author}{\bibfnamefont{D.~L.} \bibnamefont{Moehring}},
  \bibinfo{author}{\bibfnamefont{S.}~\bibnamefont{Olmschenk}},
  \bibnamefont{and} \bibinfo{author}{\bibfnamefont{C.}~\bibnamefont{Monroe}},
  \bibinfo{journal}{Phys. Rev. Lett.} \textbf{\bibinfo{volume}{100}},
  \bibinfo{pages}{150404} (\bibinfo{year}{2008}).

\bibitem[{\citenamefont{Madsen et~al.}(2006)\citenamefont{Madsen, Moehring,
  Maunz, {Kohn, Jr.}, Duan, and Monroe}}]{madsen:ultrafast-rabi}
\bibinfo{author}{\bibfnamefont{M.~J.} \bibnamefont{Madsen}},
  \bibinfo{author}{\bibfnamefont{D.~L.} \bibnamefont{Moehring}},
  \bibinfo{author}{\bibfnamefont{P.}~\bibnamefont{Maunz}},
  \bibinfo{author}{\bibfnamefont{R.~N.} \bibnamefont{{Kohn, Jr.}}},
  \bibinfo{author}{\bibfnamefont{L.-M.} \bibnamefont{Duan}}, \bibnamefont{and}
  \bibinfo{author}{\bibfnamefont{C.}~\bibnamefont{Monroe}},
  \bibinfo{journal}{Phys. Rev. Lett.} \textbf{\bibinfo{volume}{97}},
  \bibinfo{pages}{040505} (\bibinfo{year}{2006}).

\bibitem[{\citenamefont{Blinov et~al.}(2004)\citenamefont{Blinov, Moehring,
  Duan, and Monroe}}]{blinov:ion-photon}
\bibinfo{author}{\bibfnamefont{B.~B.} \bibnamefont{Blinov}},
  \bibinfo{author}{\bibfnamefont{D.~L.} \bibnamefont{Moehring}},
  \bibinfo{author}{\bibfnamefont{L.-M.} \bibnamefont{Duan}}, \bibnamefont{and}
  \bibinfo{author}{\bibfnamefont{C.}~\bibnamefont{Monroe}},
  \bibinfo{journal}{Nature} \textbf{\bibinfo{volume}{428}},
  \bibinfo{pages}{153} (\bibinfo{year}{2004}).

\bibitem[{\citenamefont{Hong et~al.}(1987)\citenamefont{Hong, Ou, and
  Mandel}}]{hong:HOM}
\bibinfo{author}{\bibfnamefont{C.~K.} \bibnamefont{Hong}},
  \bibinfo{author}{\bibfnamefont{Z.~Y.} \bibnamefont{Ou}}, \bibnamefont{and}
  \bibinfo{author}{\bibfnamefont{L.}~\bibnamefont{Mandel}},
  \bibinfo{journal}{Phys. Rev. Lett.} \textbf{\bibinfo{volume}{59}},
  \bibinfo{pages}{2044} (\bibinfo{year}{1987}).

\bibitem[{\citenamefont{Shih and Alley}(1988)}]{shih:HOM}
\bibinfo{author}{\bibfnamefont{Y.~H.} \bibnamefont{Shih}} \bibnamefont{and}
  \bibinfo{author}{\bibfnamefont{C.~O.} \bibnamefont{Alley}},
  \bibinfo{journal}{Phys. Rev. Lett.} \textbf{\bibinfo{volume}{61}},
  \bibinfo{pages}{2921} (\bibinfo{year}{1988}).

\bibitem[{\citenamefont{Braunstein and
  Mann}(1995)}]{braunstein:bs_bell_measure}
\bibinfo{author}{\bibfnamefont{S.~L.} \bibnamefont{Braunstein}}
  \bibnamefont{and} \bibinfo{author}{\bibfnamefont{A.}~\bibnamefont{Mann}},
  \bibinfo{journal}{Phys. Rev. A} \textbf{\bibinfo{volume}{51}},
  \bibinfo{pages}{R1727} (\bibinfo{year}{1995}).

\bibitem[{\citenamefont{Simon and Irvine}(2003)}]{simon:dist-entangle}
\bibinfo{author}{\bibfnamefont{C.}~\bibnamefont{Simon}} \bibnamefont{and}
  \bibinfo{author}{\bibfnamefont{W.~T.~M.} \bibnamefont{Irvine}},
  \bibinfo{journal}{Phys. Rev. Lett.} \textbf{\bibinfo{volume}{91}},
  \bibinfo{pages}{110405} (\bibinfo{year}{2003}).

\bibitem[{\citenamefont{Altepeter et~al.}(2006)\citenamefont{Altepeter,
  Jeffrey, and Kwiat}}]{altepeter:tomography}
\bibinfo{author}{\bibfnamefont{J.~B.} \bibnamefont{Altepeter}},
  \bibinfo{author}{\bibfnamefont{E.~R.} \bibnamefont{Jeffrey}},
  \bibnamefont{and} \bibinfo{author}{\bibfnamefont{P.~G.} \bibnamefont{Kwiat}},
  \bibinfo{journal}{Adv. At. Mol. Opt. Phys.} \textbf{\bibinfo{volume}{52}},
  \bibinfo{pages}{105} (\bibinfo{year}{2006}).

\bibitem[{\citenamefont{{O'Brien} et~al.}(2004)\citenamefont{{O'Brien}, Pryde,
  Gilchrist, James, Langford, Ralph, and White}}]{obrien:process_tomography}
\bibinfo{author}{\bibfnamefont{J.~L.} \bibnamefont{{O'Brien}}},
  \bibinfo{author}{\bibfnamefont{G.~J.} \bibnamefont{Pryde}},
  \bibinfo{author}{\bibfnamefont{A.}~\bibnamefont{Gilchrist}},
  \bibinfo{author}{\bibfnamefont{D.~F.~V.} \bibnamefont{James}},
  \bibinfo{author}{\bibfnamefont{N.~K.} \bibnamefont{Langford}},
  \bibinfo{author}{\bibfnamefont{T.~C.} \bibnamefont{Ralph}}, \bibnamefont{and}
  \bibinfo{author}{\bibfnamefont{A.~G.} \bibnamefont{White}},
  \bibinfo{journal}{Phys. Rev. Lett.} \textbf{\bibinfo{volume}{93}},
  \bibinfo{pages}{080502} (\bibinfo{year}{2004}).

\bibitem[{\citenamefont{Horodecki et~al.}(1999)\citenamefont{Horodecki,
  Horodecki, and Horodecki}}]{horodecki:teleport_channel}
\bibinfo{author}{\bibfnamefont{M.}~\bibnamefont{Horodecki}},
  \bibinfo{author}{\bibfnamefont{P.}~\bibnamefont{Horodecki}},
  \bibnamefont{and}
  \bibinfo{author}{\bibfnamefont{R.}~\bibnamefont{Horodecki}},
  \bibinfo{journal}{Phys. Rev. A} \textbf{\bibinfo{volume}{60}},
  \bibinfo{pages}{1888} (\bibinfo{year}{1999}).

\end{thebibliography}

\end{document}